\documentclass[aps,superscriptaddress,preprintnumbers,
showpacs,a4,twoside,twocolumn, prl]{revtex4}

\usepackage{bm}
\usepackage{graphicx}
\usepackage{amsfonts}
\usepackage{amsmath}
\usepackage{amssymb}

\begin{document}

\title{Universal phase diagram of disordered bosons from a doped quantum 
magnet}

\author{Rong Yu}
\affiliation{Department of Physics \& Astronomy, Rice University, Houston, TX 77005, USA}     
\author{Stephan Haas}
\affiliation{Department of Physics and Astronomy, University of Southern 
California, Los Angeles, CA 90089-0484, USA}
\author{Tommaso Roscilde}
\affiliation{Laboratoire de Physique, Ecole Normale Sup\'erieure de Lyon,
46 All\'ee d'Italie, 69007 Lyon, France}
\affiliation{Max-Planck-Institut f\"ur Quantenoptik, Hans-Kopfermann-strasse 1,
85748 Garching, Germany}

\begin{abstract} 
A quantum particle cannot in general diffuse through a disordered
medium because of its wavelike nature, but interacting particles can
escape localization by collectively percolating through 
the system. For bosonic particles this phenomenon corresponds to a quantum 
transition from a localized insulator phase -  the Bose glass - to a 
superfluid phase, in which particles condense into an extended state. 
Here, we construct a universal phase diagram of disordered bosons  
in doped quantum magnets for which bosonic quasi-particles are represented by 
magnetized states (spin triplets) of the quantum spins, condensing into a magnetically
ordered state. The appearance of a Bose glass leads to strong measurable signatures 
in the onset of superfluidity of the spin-triplet gas, exhibiting a complex crossover from 
low-temperature quantum percolation to a conventional 
condensation transition at intermediate temperatures.
\end{abstract}
\maketitle

In continuous quantum phase transitions \cite{Sondhietal97} quantum 
fluctuations drive a complete change of symmetry in the state of the 
system via the development of long-wavelength modes at arbitrarily low 
energies. This occurs in the case of strongly interacting bosons on a lattice,
modeling widely different physical systems such as cold 
atoms loaded in optical lattices \cite{Greineretal02}, Cooper pairs in 
Josephson-junction arrays \cite{Mooijetal96}, or magnetic quasi-particles 
in gapped quantum magnets \cite{Giamarchietal07}. In the presence of 
strong repulsion among the particles, a commensurate filling of the 
lattice stabilizes a gapped Mott insulating phase, which has all the 
symmetries of the original Hamiltonian. In contrast, doping the system 
away from commensurate filling drives it into a Bose-condensed phase 
\cite{Fisheretal89}, characterized by superfluidiy and a gapless 
phonon-like mode.

The presence of disorder in the system can deeply affect its quantum 
critical behavior. In fact, disorder can localize the long-wavelength modes developed
at the quantum phase transition of the clean system, and thus introduce
a novel phase, with gapless localized excitations 
but in the absence of any long-range order. Indeed, it has been shown that  
lattice disorder for interacting bosons can localize the gas of doped 
particles/holes, hindering their condensation and thus giving rise to a 
new insulating gapless phase with incommensurate filling, known as a 
\emph{Bose glass} \cite{GiamarchiS88,Fisheretal89}. Moreover, when the 
density of doped particles/holes exceeds a critical threshold, 
their mutual interactions destabilize energetically the localized state 
in favor of a delocalized  
superfluid state. It is clear that the nature of this transition is completely 
different from that of the Mott-insulator/superfluid transition 
in the absence of randomness. 
 
 Despite its extreme richness and generality, the phase diagram of 
so-called dirty-boson systems has so far been extremely elusive, both
to experiments, due to the difficulty of accessing
the quantum critical regime in a controlled disordered environment  
\cite{Fallanietal07, Reppyetal, Oudenaarden96}; and to theory, due to the absence of well-controlled 
approximations in presence of strong interactions and strong disorder
\cite{Weichmann}. Here we propose the use of doped quantum magnets 
to fully investigate the physics of dirty bosons, and we perform  
numerically exact simulations on a realistic model Hamiltonian to 
construct the universal phase diagram of the system. Our model of interest
is the magnetic Hamiltonian of NiCl$_2$$\cdot$4SC(NH$_2$)$_2$ (DTN) 
\cite{Zapfetal06,Zvyaginetal07}, which consists of coupled $S=1$
chains with strong single-ion anisotropy:
\begin{eqnarray}
{\cal H} &=& J_{c} \sum_{\langle ij \rangle_{c}} \epsilon_i \epsilon_j 
{\bm S}_{i}\cdot{\bm S}_{j} ~~+ 
J_{ab} \sum_{\langle lm \rangle_{ab}} \epsilon_l \epsilon_m
{\bm S}_{l}\cdot{\bm S}_{m} \nonumber \\
&+& D \sum_{i} \epsilon_i (S^z_{i})^2
- g\mu_B H \sum_{i} \epsilon_i S^z_{i}.
\label{e.Ham}
\end{eqnarray}
where $J_{c}>0$ is the dominant antiferromagnetic coupling for bonds
$\langle ij \rangle_{c}$ along the 
crystallographic $c$-axis, $J_{ab}>0$ is the coupling for bonds 
$\langle lm \rangle_{ab}$ in the transverse
$a$, $b$ directions of the cubic lattice \cite{notecrystal}, and $D>0$ 
is the single-ion anisotropy. We adopt the experimental parameters
\cite{Zapfetal06,Zvyaginetal07} $J_{c}=2.2$ K , $J_{ab}=0.18$ K, $D=8.9$ K.
In the following we will use reduced temperatures $t = k_B T/J_c$
and reduced magnetic fields $h= g\mu_B H/J_c$.
Moreover we introduce randomness in the system in the form
of site dilution via the variables $\epsilon_i$ taking values 0 and 1
with probability $x$ and $1-x$ respectively, where $x$ is the dilution
concentration. Site dilution corresponds experimentally to Mg$^{2+}$ 
or Cd$^{2+}$ doping of the Ni$^{2+}$ ions.

\begin{figure}[h]
\begin{center}
\includegraphics[
     width=75mm,angle=0]{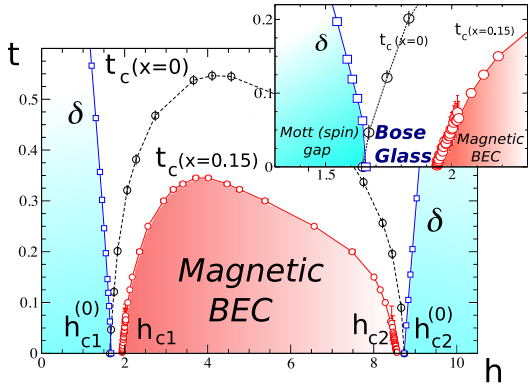} 
 \null\vspace*{-.5cm}
\caption{Phase diagram of coupled anisotropic $S=1$ chains 
in a magnetic field with 15\% dilution.  We plot the critical temperature 
$t_c$ for magnetic BEC and the spin gap
$\delta=\Delta/J_c$  \cite{spin-gap-estimate} as a function of the 
applied magnetic field for
zero doping ($x=0$) and for 15\% doping ($x=0.15$). The spin gap
is independent of the doping concentration. The inset shows a
zoom on the region close to the lower critical field $h_{c1}$,
marking the appearance of the Bose glass.}
\label{f.phdiagr}
\end{center}
\null\vspace*{-1.2cm}
\end{figure}

\begin{figure}[h]
\begin{center}
\includegraphics[
     width=75mm,angle=0]{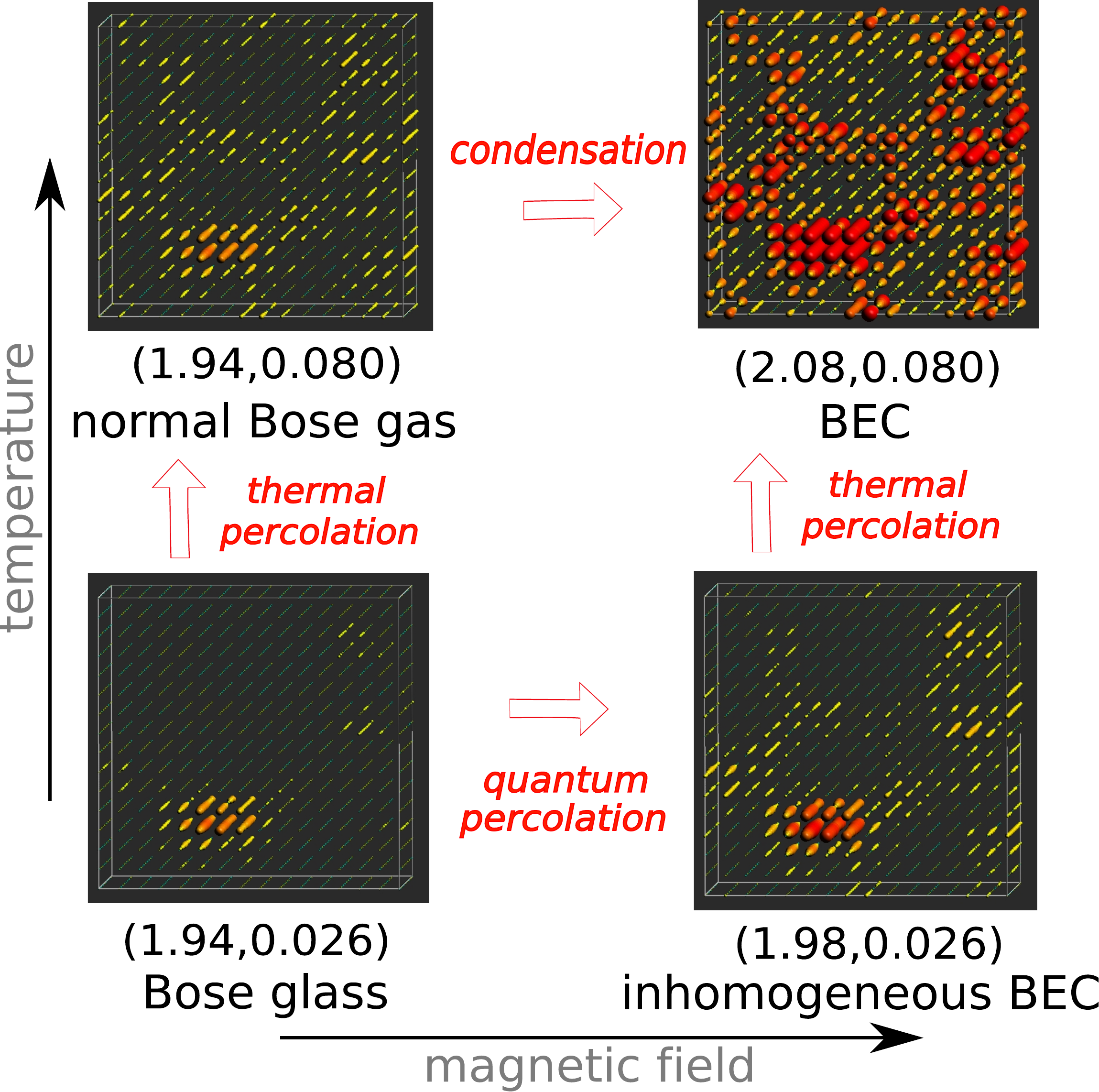} 
\caption{ Local magnetization profile (corresponding to
the excess spin-triplet bosons) of coupled $S=1$ chains, Eq.~\eqref{e.Ham}
on a $16^3$ lattice with 15$\%$ dilution, obtained via quantum 
Monte Carlo simulations. Four representative points in the $h$ -- $t$
plane are shown  [$(h,t)$ coordinates are indicated below each figure],
for the following bosonic phases: (1) a \emph{Bose glass} with localized incoherent 
excess spin quasiparticles; (2) an \emph{inhomogeneous BEC phase}, 
with long-range coherence due to weak links between localizes states; (3) 
a \emph{normal Bose gas}, with incoherent spin quasiparticles which are
spread over a percolating region; (4) a \emph{BEC} phase, with long-range
coherence of spin quasiparticles homogeneously spread over a percolating 
region. In addition to these phases, the \emph{Mott insulator} (not shown), 
is characterized by absence of 
excess spin quasiparticles. The pictures shown here refer to the regime
$h \sim h_{c1}$.  An analogous succession of phases, mirrored along 
the magnetic field axis, is observed for $h \sim h_{c2}$.}
\label{f.phases}
\end{center}
\null\vspace*{-1.2cm}
\end{figure}

 A standard spin-to-boson mapping \cite{Auerbach94} allows 
to recast the above Hamiltonian in the form of bosonic
operators $b_i$, $b_i^{\dagger}$,
with the local Hilbert space limited to the Fock states $|n\rangle=
|0\rangle$, $|1\rangle$ and $|2\rangle$ corresponding to 
the states $|m_S\rangle$= $|-1\rangle$, $|0\rangle$ and $|1\rangle$ 
of a $S=1$ spin. Upon this transformation the $J$ couplings take the
role of nearest-neighbor hopping amplitudes and repulsion potentials,
while the single-ion anisotropy plays the role of 
on-site repulsion for the bosons, and finally the magnetic field
acts as the chemical potential. 

 Due to the large anisotropy $D$, the ground state
of the system in zero field is well approximated by a collection of 
singlets
$|m_S=0\rangle$, corresponding to a gapped Mott insulator of bosons with
$n=1$ particle per site. In the case of the pure system \cite{Zapfetal06}, 
applying a magnetic 
field above a critical value $h_{c1}^{(0)}= 1.65(1)$  allows to close the gap in 
the spectrum and to increase the particle density above unit filling, 
corresponding to the appearance of a finite magnetization along the field 
direction. Upon increasing the filling, the system develops a superfluid and 
condensate fraction, 
which translates into the spin stiffness of the magnetic system and into a peak
at $\bm q = (\pi,\pi,\pi)$ in the transverse static structure factor,
$S_{\perp}(\bm q) = \sum_{ij} \exp[i{\bm q}\cdot({\bm r}_i - {\bm r}_j)] \langle 
S_i^{x(y)}  S_j^{x(y)} \rangle /[(1-x)L^3]$, 
respectively. The condensation order parameter is directly
related to the spontaneous magnetization appearing in the $xy$ plane 
\cite{magneticBEC}, 
$\langle b_i \rangle \approx \langle S_i^x - iS_i^y \rangle/\langle \sqrt{S- S_i^z}\rangle $.
Condensation (in the form of magnetic order)
survives at finite temperature up to a critical temperature $t_c$, which
scales with the applied field as $t_c \sim (h-h_{c1}^{(0)})^{\phi_0}$,
where $\phi_0=2/3$ as predicted by mean-field theory for the condensation
of a diluted Bose gas \cite{PethickS02}. The experimental verification of 
this prediction \cite{magneticBEC} is considered as the strongest evidence
for magnetic Bose-Einstein condensation (BEC). Increasing the field even further
reduces the state of the system to that of a dilute gas of singly occupied sites 
(excess triplet holes), associated with $|m_s = 0\rangle$ spin states. A high field
eventually leads to full polarization of the spins at the critical value $h_{c2}^{(0)}=8.69$, 
corresponding  to a second Mott insulating state of triplets (namely the vacuum of
triplet holes), and to a vanishing of the 
critical temperature which obey the same scaling law, $t_c \sim (h_{c2}^{(0)}-h)^{\phi_0}$.


In the presence of site dilution the phase diagram changes substantially.
 The results of an extensive quantum Monte Carlo study \cite{method} 
 for a dilution of $x=$15\%  are shown in Fig.~\ref{f.phdiagr}. 
 The most important 
 effect of disorder on the phase diagram is that
of shifting the critical fields for condensation and shrinking the 
condensation region with respect to the clean system,  
$h_{c1} \approx 1.94(1) > h_{c1}^{(0)}$ and $h_{c2} \approx 8.52(2)  <  h_{c2}^{(0)}$.
This effect separates the onset of
condensation from the closing of the spin gap, which still occurs 
at $h_{c1}^{(0)}$ and $h_{c2}^{(0)}$. In fact, the spin gap of the doped 
system is the \emph{same}
as for the clean system, given that the minimum spin gap 
in the doped system is associated with local excitations in rare regions 
that locally approximate the clean system, and whose size can be arbitrarily
large. Hence disorder ``splits" an
ordinary quantum critical point into two, i.e. a spectral transition point 
for 
the closing of the gap, and a true critical point for the onset of order.
Inbetween these two critical points an extended Bose-glass phase
opens up, characterized by the joint absence of a spin gap \emph{and} of 
superfluidity. In the magnetic language, this represents a gapless quantum 
paramagnetic phase. 
In the clean case of the Hamiltonian given in 
Eq.~\eqref{e.Ham} the $T=0$ quantum-critical point is Gaussian, 
with mean-field critical exponents and dynamical critical exponent
$z=2$ \cite{Fisheretal89}. In particular, the spin gap closes as
$\delta\sim |h-h_{c1(2)}^{(0)}|^{\nu z} = |h-h_{c1(2)}^{(0)}|$, where
$\nu=1/2$ is the mean-field correlation length exponent. In the 
disordered system the gap still closes in a mean-field-like fashion, but 
disorder completely discards the mean-field 
picture for the true quantum-critical points at $h_{c1}$ and 
$h_{c2}$, which 
obey the exponents of a \emph{new} universality class pertinent to the
superfluid-to-Bose-glass transition. The prediction of 
Ref.~\onlinecite{Fisheretal89} for the dynamical critical exponent, 
$z=d=3$, is well verified by our numerical data. Moreover, we 
determine the correlation length exponents as $\nu = 0.7(1)$, 
consistent with Ref.~\onlinecite{Sorensen06}, and the order-parameter
exponent as $\beta = 0.9(1)$ \cite{Yutoappear}.

\begin{figure}
 \begin{center}
   \includegraphics[width=7.5cm]{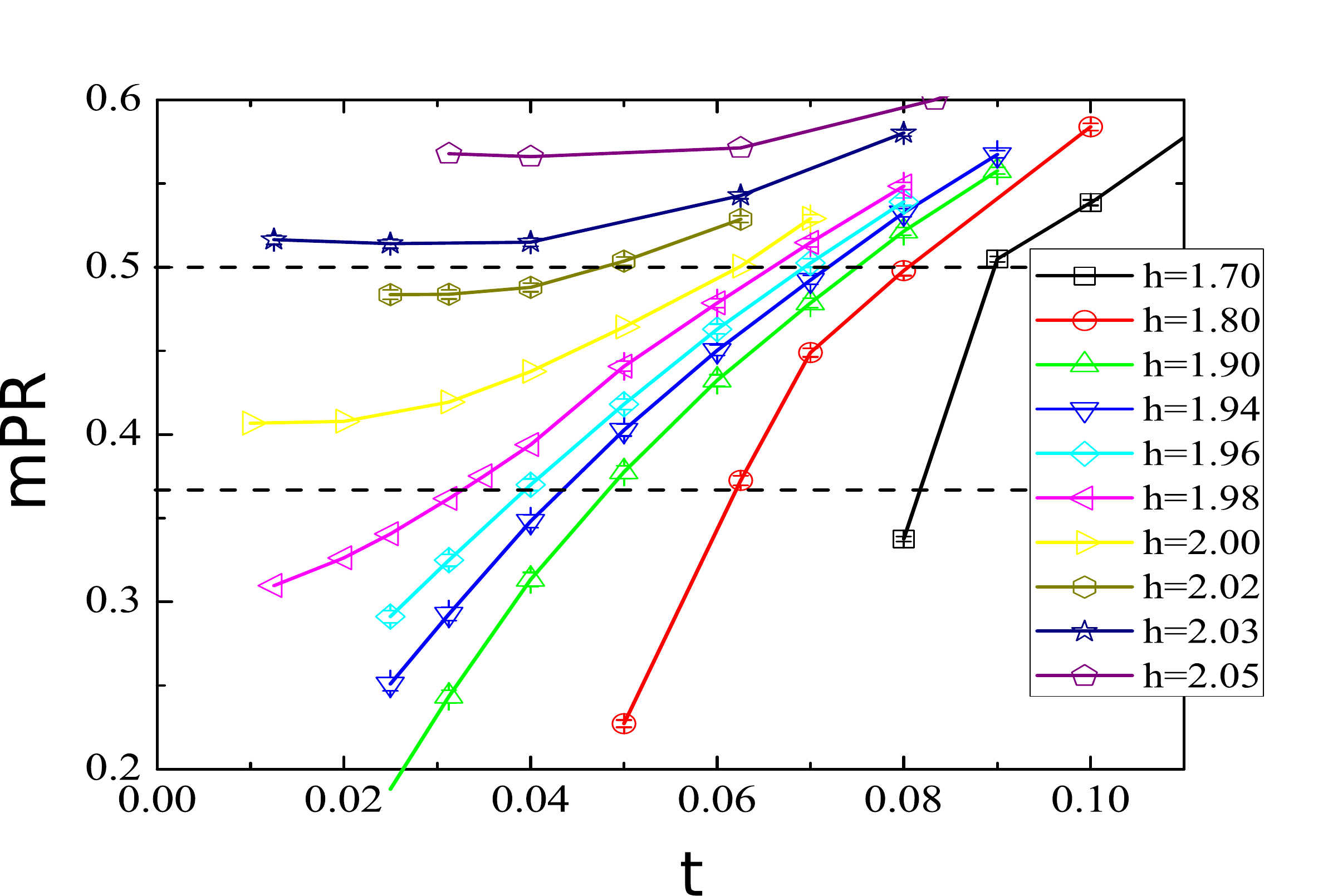} \\
   \caption{ 
   Magnetic participation ratio (see definition in the text) as a function of temperature 
   for different magnetic fields. The two thresholds mPR$=0.366$  and mPR$=0.5$ 
   are indicated by dashed lines.}
   \label{f.percola}
   \end{center}
   \null\vspace*{-1.2cm}
   \end{figure}

 The Bose-glass state is characterized by Anderson localization \cite{Anderson58}
 of the excess spin triplets (or triplet holes) injected into the Mott insulating state: this
 leads to a magnetization profile characterized by magnetized
``puddles", located in the rare regions devoid of lattice vacancies,
 as shown by direct inspection of the numerical data (see Fig.~\ref{f.phases}). 
To characterize the density profile of the localized spin triplets (triplet holes)
we investigate the behavior of the magnetic participation ratio (mPR)  
\begin{equation}
   {\rm mPR} = \frac{1}{(1-x)L^3} \left \langle
   \frac{\left(\sum_i \langle S_i^z \rangle \right)^2}{\sum_i \langle S_i^z \rangle^2}
   \right\rangle_{\rm disorder}
   \end{equation}
representing the effective fraction of the system which is magnetized,
Here $\langle...\rangle_{\rm disorder}$ represents the disorder average. 
Fig.~\ref{f.percola} show the mPR as a function of temperature for various magnetic
fields. A percolating
magnetization profile is characterized by a participation ratio which 
exceeds a fraction $p^*(x)=p_{\rm sc}/(1-x)=0.366$, corresponding
to the percolation threshold for the simple cubic lattice $p_{\rm sc}=0.312$ \cite{StaufferA94}
renormalized by the dilution.
Below this threshold, the magnetization profile is composed of spatially
disconnected puddles, which are phase-incoherent, resulting in the absence of 
long-range order of the transverse spin components. 
This remains true for
a significant portion of the finite-temperature region above the 
zero-temperature
Bose glass, despite the strong thermal activation of spin triplets/holes. 
We term this temperature region as \emph{thermal Bose glass}. 
In such a region the magnetization 
$m(t,h) =  \sum_i{\langle S_i^z \rangle} / [(1-x)L^3]$ 
can be well reproduced by a model of disconnected clusters with sizes obeying
the statistics of rare clean regions in a site-diluted cubic lattice
\cite{Yutoappear}.

 Increasing the temperature above the characteristic value
 $t_{{\rm mPR}=p^*}$ at which the magnetic participation ratio reaches the 
 condition mPR $= p^*$, one enters into a thermal percolation
 crossover region, marking the upper bound of the thermal 
 Bose glass. Above this region the spin triplets / holes, albeit
 incoherent, are delocalized over the entire lattice. Thermal
 percolation is marked by a substantial change in the statistics
 of the local magnetization $\langle S_i^{z} \rangle$, corresponding
 to the local excess population of triplets / holes. 
 Indeed one can easily show that mPR $= [1+(\sigma/m)^2]^{-1}$,
 where $m$ is the average of the $\langle S_i^{z} \rangle$
 distribution (namely the total magnetization) and $\sigma$ its standard
 deviation. In particular the link between mPR and the local 
 magnetization distribution suggests a second criterion for
 thermal percolation. Considering the ratio
 $R=m/\sigma$, we take $R=1$ (corresponding
 to mPR$=1/2$) as the crossover value  above which 
 the fluctuations are dominated by the mean. This crossover
 marks the passage from a highly inhomogeneous magnetization
 profile to a homogeneous one. 
 Hence in the following we adopt the two criteria (mPR$= p^*$
 and mPR$=1/2$) to mark the thermal percolation crossover. 
 Most importantly, in absence of spontaneous ordering transverse to the field, 
the $\langle S_i^{z} \rangle$ distribution is probed  
by the lineshape of nuclear magnetic resonance \cite{Tedoldietal99}, 
which means that thermal percolation can be experimentally
detected.  
 
  Given the above analysis, it is to be expected 
 that the transition to triplet/hole condensation, induced by the applied field, 
will be very different when coming from the thermal Bose glass
or when coming form the normal Bose gas. In the former case, 
indeed, the triplet/hole gas undergoes percolation of quantum coherence,
and hence condensation, via tunneling of the triplet/hole bosons between 
the magnetization puddles; 
this clearly represents a finite-temperature \emph{quantum percolation} 
phenomenon. In the latter case, in contrast, condensation 
occurs on the magnetized backbone of the diluted lattice, akin
to field-induced condensation in a clean system.  
 
\begin{figure}
 \begin{center}
   \includegraphics[width=6.5cm]{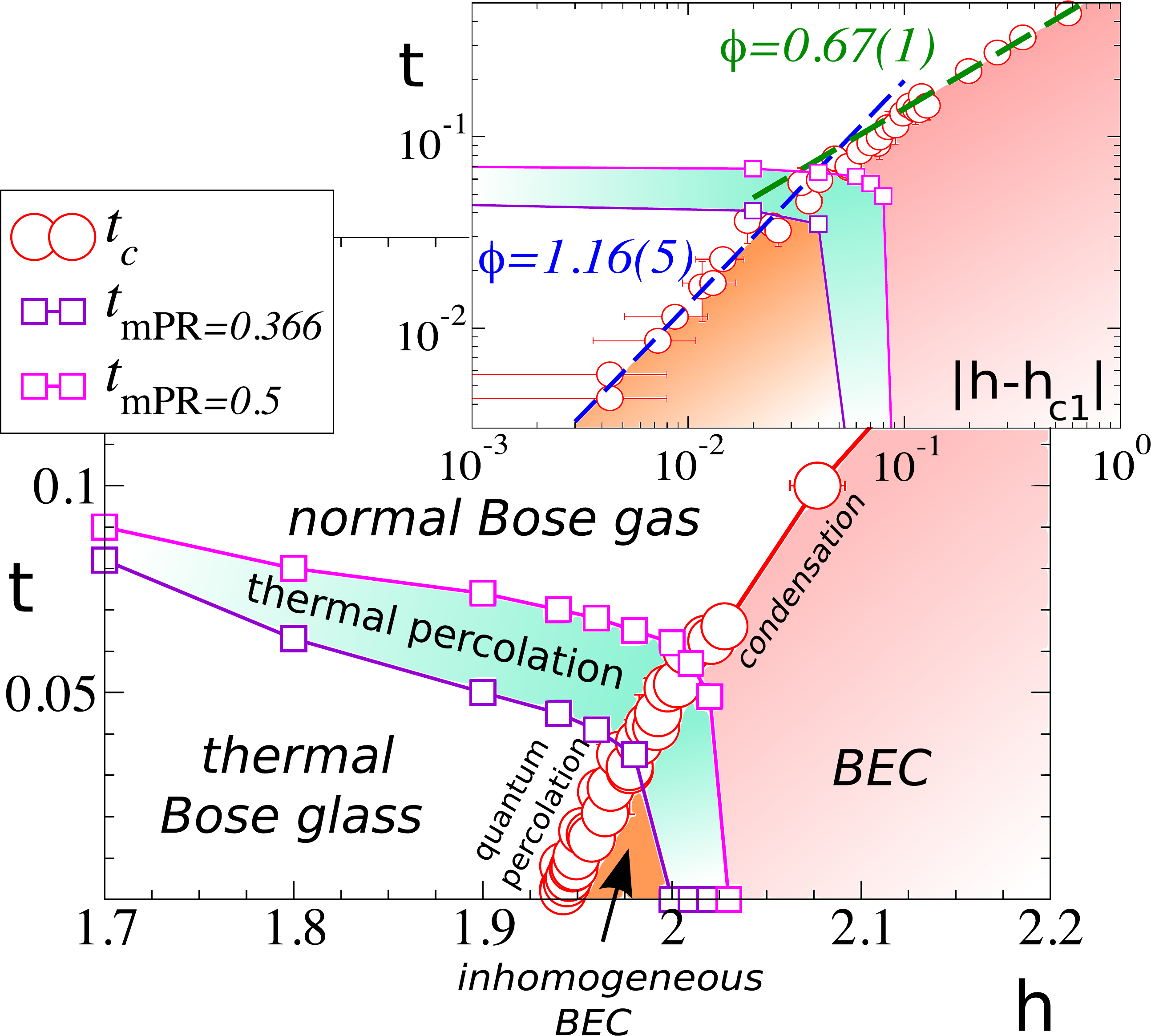}~~~~~~\\ 
   \caption{Percolation phase diagram of site-diluted coupled
   $S=1$ chains in a magnetic field. In the inset: log-log plot
   of $t_c$ vs $|h-h_{c1}|$.  A completely analogous behavior
   is observed close to the upper critical field $h_{c2}$.}
   \label{f.percolation-PhD}
   \end{center}
   \null\vspace*{-.9cm}
   \end{figure}

  As shown in Fig.~\ref{f.percolation-PhD}, this crossover 
  from low-temperature quantum percolation 
 to more conventional condensation on a random network has a 
 striking signature in the central feature of magnetic BEC, 
 namely the scaling of the critical temperature close to the 
 lower critical field. The Bose-glass nature of the disordered
 phase at low temperature leads to the breakdown of the 
 scaling prediction stemming from mean-field theory of 
 BEC in a dilute Bose gas. Numerically we find that the 
 onset of $T_c$ follows  $t_c \propto |h - h_{c1(2)}|^\phi$ 
 with $\phi=1.16(5)$, in violation of the mean-field prediction
  \cite{Yutoappear}.
  But, even more remarkably, above
  a characteristic temperature range the 
  $t_c(h)$ curve shows a clear \emph{crossover} to a 
  different curvature, consistent this time with the mean-field
  exponent $\phi_0=2/3$. We observe that this crossover 
  corresponds to the dividing 
  region between the two different condensation regimes, namely 
 from a thermal Bose glass and from a normal Bose gas. 
 Indeed, above the crossover region spatial
 percolation of the magnetization profile occurs already 
 before condensation, and hence condensation 
 acquires a conventional character, while disorder only
 plays the role of effectively renormalizing the Hamiltonian
 parameters of the clean system. It is important to stress that
 the disorder-induced crossover in the scaling behavior
 is \emph{not} accompanied by a change in the universality class
 of the transition: indeed,  according to the Harris criterion
 \cite{Harris} disorder is not pertinent for the 3D XY 
 condensation transition of the clean system, for which
$\nu \approx 0.67 \geq 2/D = 2/3$ \cite{PelissettoV02}.
 
 In conclusion, we have shown that a realistic disordered quantum spin model,
 corresponding to the magnetic Hamiltonian of DTN \cite{Zapfetal06}
 under non-magnetic doping, can elucidate the universal features
 of the physics of strongly correlated bosons on a disordered lattice in the
 grand-canonical ensemble. A variety of measurement tools, such
 as magnetometry and NMR, can reveal the central features of 
 Anderson localization of triplet bosons, leading to inhomogeneous 
 magnetization profiles, and to thermal and/or quantum 
 percolation thereof. These results clearly demonstrate the potential of model 
 quantum magnets as a valuable testbed for theories of complex Bose 
 systems with strong interactions and disorder, and more generally 
 for theories of the fundamental interplay between quantum critical fluctuations
 and geometric randomness.  
 
The authors wish to acknowledge V. Zapf, L. Yin, A. Paduan-Filho, and M. Jaime for 
stimulating discussions. S. H. is supported by DOE grant DE-FG02-06ER46319. 
The numerical simulations have been performed on
the computer facilities of the NCCS  at the Oak Ridge National Laboratories, 
and supported by the INCITE Award MAT013 of the Office of 
Science - U.S. Department of Energy.


\begin{thebibliography}{99}
\bibitem{notecrystal} The structure of DTN is actually that of 
two interpenetrating tetragonal lattices, which can be considered 
effectively as decoupled \cite{Zvyaginetal08}.
\bibitem{spin-gap-estimate} The spin gap $\delta$ has been estimated by
fitting the low-temperature behavior of the uniform susceptibility $\chi(t\to0)$
to the exponential form $B \exp(-\delta/t)$.
\bibitem{method}
The phase diagram of the disordered quantum spin system
has been obtained via numerically exact quantum Monte Carlo \cite{SSE} 
on lattice sizes as large as $16^3$, and averaging all results over 
$\sim 300$ disorder realizations. The critical temperatures, critical fields
and critical exponents have been estimated via a careful finite-size scaling analysis
of the spin structure factor, correlation length and spin stiffness \cite{Yutoappear}.
\end{thebibliography}
\end{document}